\documentclass[aip, apl, reprint]{revtex4-1}
\usepackage{graphicx}

\newcommand{\degree}{{$^{\rm o}$}}

\newcommand{\kv}{{\bf k}}

\begin{document}

\title{The Coherent Interlayer Resistance of a Single, Rotated
Interface between Two Stacks of AB Graphite}

\author{K. M. Masum Habib}
\email[Email:~]{khabib@ee.ucr.edu}
\affiliation{Department of Electrical Engineering, University of
California, Riverside, CA 92521-0204}
\author{Somaia S. Sylvia}
\affiliation{Department of Electrical Engineering, University of
California, Riverside, CA 92521-0204}
\author{Supeng Ge}
\affiliation{Department of Physics and Astronomy, University of
California, Riverside, CA 92521-0204}
\author{Mahesh Neupane}
\affiliation{Department of Electrical Engineering, University of
California, Riverside, CA 92521-0204}
\author{Roger K. Lake}
\email[Email:~]{rlake@ee.ucr.edu}
\affiliation{Department of Electrical Engineering, University of
California, Riverside, CA 92521-0204}

\begin{abstract}
The coherent, interlayer resistance of a misoriented, 
rotated interface 
between two stacks of AB graphite
is determined for a variety of misorientation angles. 
The quantum-resistance of the ideal AB stack is on the order 
of 1 to 10 m$\Omega \mu{\rm m}^2$.
For small rotation angles, 
the coherent interlayer resistance 
exponentially approaches the ideal quantum resistance 
at energies away from the charge neutrality point.
Over a range of intermediate angles, the resistance increases 
exponentially with cell size for minimum size unit cells. 
Larger cell sizes, of similar angles, may not follow this trend. 
The energy dependence of the interlayer transmission is described.
\end{abstract}


\maketitle 

There is rapidly growing interest in vertically stacked van der Waals materials
for electronic device applications.\cite{Britnell_graphene_BN_vFET_Sci12,
Barristor_Sc12,
Javey_InAs_WSe2_diode_APL13,
Vertical_FET_Geim_Novoselov_Nat12,
Vert_stack_inverters_NatMat13,
Vert_graph_base_EDL12,
Jena_ProcIEEE13}
In such structures the interfaces between different materials
will, in general, be misoriented with respect to each 
other.\cite{Geim_Grigorieva_review_Nat13}
THz cutoff frequencies 
have been predicted for such devices.\cite{Vert_graph_base_EDL12}
At such high frequencies, any small series resistance can degrade performance.
For example, an emitter contact resistance of 2.5 $\Omega \mu{\rm m}^2$
is required to achieve a THz cutoff frequency in a heterostructure
bipolar transistor.\cite{Rodwell_ProcIEEE08}
Understanding the effect of the misorientation on the interlayer resistance
is required to fully understand the design requirements and
performance of proposed vertically stacked devices.

The most well studied and well understood of the
van der Waals material are
graphite and graphene.
\cite{Morgan_Uher_PhilMag81,Dresselhaus_c-axis_R_Carbon76,S.Ono_c-axis_R_JPSJ76,Neto_RevModPhys09,Geim_Grigorieva_review_Nat13}
There is a long history of 
investigations of the \textit{c}-axis resistance of 
graphite\cite{Primak_Fuchs_PRB54,
Dresselhaus_c-axis_R_Carbon76,
S.Ono_c-axis_R_JPSJ76,
Tsuzuku_graphite_review83,
Tsuzuku_c-axis_R_PRB90,
Nano_stacked_jns_JJAP11}
This body of work focused on stacks of kish graphite or 
highly ordered pyrolytic graphite
with a random ensemble of stacking faults in the diffusive limit.

The effect of misorientation on the electronic structure of bilayer
graphene has been studied extensively both theoretically and 
experimentally.\cite{Santos_band_twisted_PRL07,
Wu_chiral_electrons_PRB07,
Latil_DFT_twisted_PRB07,
Hass_twisted_SiC_PRL08,
Shallcross_bi_twist_PRL08,
Shallcross_band_turbo_graphene_PRB10,
Laissardiere_e_localization_NL10,
Luican_bi_twisted_PRL11,
Santos_cont_model_PRB12,
Shallcross_PRB13}
After a few degrees misorientation, the in-plane dispersion becomes linear,
and after about 10 degrees misorientation, the in-plane velocity is
the same as that of single-layer graphene.
Thus, the two misorineted layers of 
graphene act as if they are electronically decoupled.

The interlayer resistance of misoriented bilayer graphene 
has received less 
attention.\cite{Bistritzer_transport_twisted_PRB10,Avouris_twisted_PRL12,Kim_coherence_PRL13}
The calculated coherent interlayer resistance as a function 
of rotation angle $\theta$
is found to vary by 16 orders of magnitude as 
the misorientation angle changes
from zero to 30 degrees.\cite{Bistritzer_transport_twisted_PRB10}
The values vary from 
approximately $10^{15}$ $\Omega \mu{\rm m}^2$
to $0.1$ $\Omega \mu{\rm m}^2$.
The room-temperature, phonon-mediated interlayer resistance 
of misoriented bilayer graphene
shows far less dependence on the misorientation 
angle.\cite{Avouris_twisted_PRL12,Kim_coherence_PRL13} 
It changes by
less than an order of magnitude as the angle varies from zero to 30 
degrees.\cite{Avouris_twisted_PRL12,Kim_coherence_PRL13}
Its calculated value is approximately 100 $\Omega \mu{\rm m}^2$ over a range
of intermediate rotation angles.\cite{Avouris_twisted_PRL12}
Experimental measurements found approximately
an order of magnitude larger resistance that varied 
from 750 $\Omega \mu{\rm m}^2$
to 3400 $\Omega \mu{\rm m}^2$ as the angle varied from
$5^\circ$ to $24^\circ$.\cite{Kim_coherence_PRL13,Kim_coherence_PRL13_note}
Calculations of the interlayer magnetoresistance of misoriented bilayer
graphene ribbons show a large magnetoresistance ratio accompanied by large transmission
peaks or Fano resonances resulting from edge states.\cite{Sonia_JAP13}
\begin{figure}
\includegraphics[width=3in]{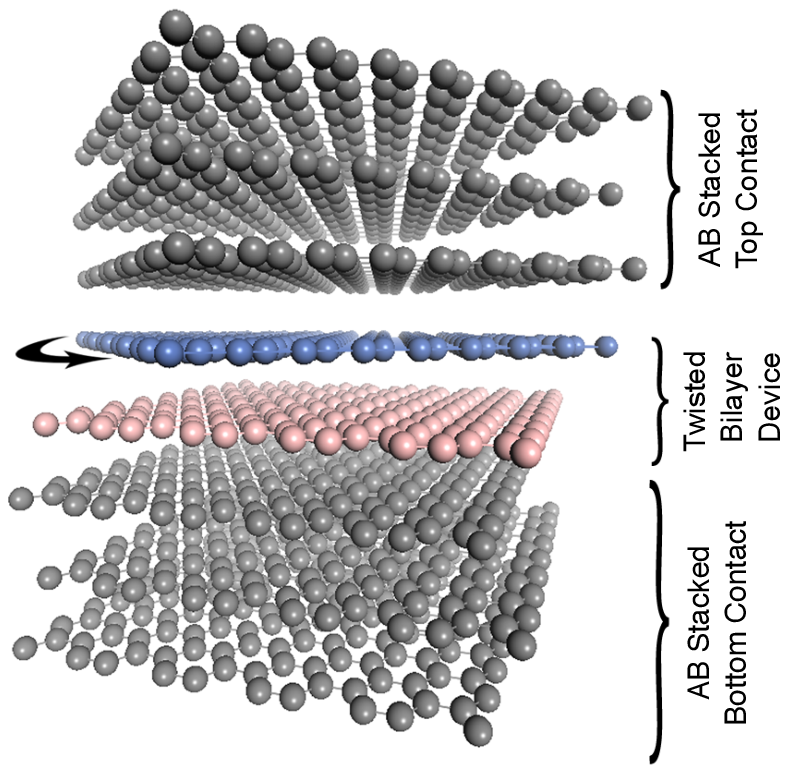}
\caption{
(Color Online) Atomistic geometry of the rotated interface. 
It consists of two AB oriented stacks that 
are rotated with respect to each other. 
The interface layers where the misorientation occurs have been
colored for visualization. The two misoriented layers are the
`device' in the NEGF calculation.
\label{fig:device}
}
\end{figure}

In this work, we calculate the transmission through
two stacks of AB graphite that are rotated
with respect to each other at their interface.
In such a structure, the semi-infinite AB graphite stacks act as
ideal leads so that
injection is well defined using the usual non-equilibrium Green 
function (NEGF) approach. 
The resistance can be calculated for $\theta = 0$\degree~providing a minimum
baseline value.
This type of structure is consistent with 
the proposed vertically stacked van der Waals structures.
We determine the coherent, interlayer resistance for
a wide range of rotation angles.
The energy dependence of the coherent interlayer
resistance is calculated and discussed.

%
The twisted bilayer graphene (TBG) supercell
(i.e. the primitive cell of the commensurate twisted bilayer) 
is created following the method described in 
Ref. [\onlinecite{Shallcross_band_turbo_graphene_PRB10}].
The top layer of the TBG supercell is used to create 
an AB stacked bilayer graphene supercell which, in turn, 
is used to create the top contact.
Similarly, the bottom contact is created using the 
bottom layer of the TBG supercell.
Thus, the twisted structure consists of
two AB oriented stacks that are rotated with
respect to each other as shown in Fig. \ref{fig:device}.
The inter layer coherent transport through the twisted structure is modeled 
using the non-equilibrium Green function (NEGF) formalism
with an empirical tight binding Hamiltonian.
The coherent resistance is calculated using
\begin{equation}
R = 1/\left[ 2\frac{\rm e^2}{\hbar}\int\frac{dE}{2\pi}~
T(E)\left(-\frac{\partial f}{\partial E}\right)\right]
\label{eq:R}
\end{equation}
where $f(E)$ is the Fermi function. The transmission $T(E)$ is 
given by
$T(E) = \int_{\rm 1^{st} BZ} d\kv~ T(E,\kv)$
where $\kv$ is 2D wave vector in the TBG Brillouin zone
and $T(E,\kv)$ is the wavevector
resolved transmission calculated using NEGF. 
A tight-binding Hamiltonian is used. 
The in-plane 
nearest neighbor hopping element is $t=3.16$ eV.\cite{Neto_RevModPhys09}
The model developed by Perebeinos et al. is used for the 
out-of-plane coupling.\cite{Avouris_twisted_PRL12}
Details of the methods are given in the Supplementary 
Information.\cite{supplement2}

\begin{figure*}
\includegraphics[width=6.5in]{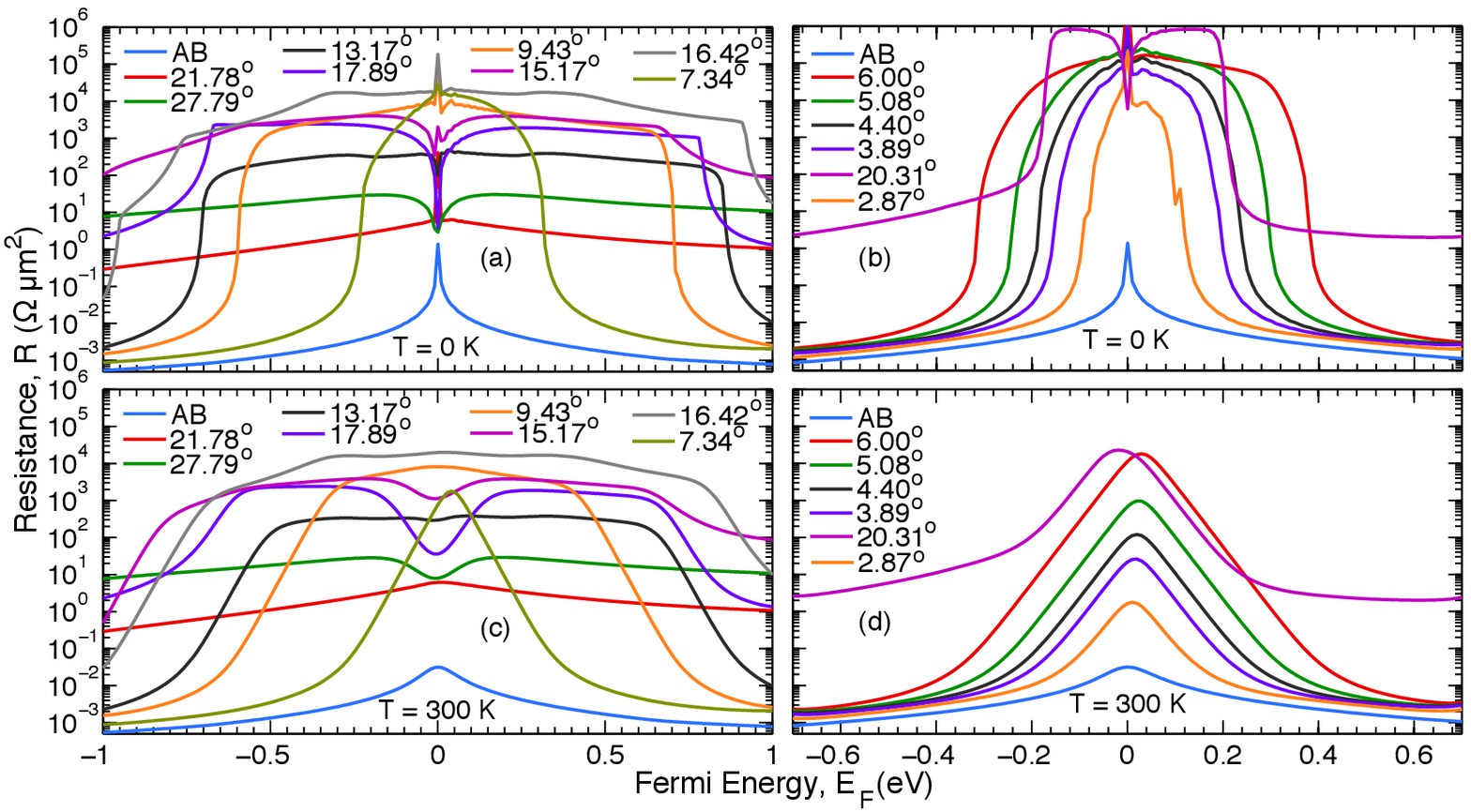}
\caption{
(Color Online) (a, b) Zero temperature coherent contact resistance 
of twisted bilayer graphene as a function of Fermi Energy 
for different rotation angles. 
(c, d) Room temperature coherent contact resistance 
of twisted bilayer graphene as a function of Fermi energy 
for different rotation angles.
\label{fig:plotRAll}
}
\end{figure*}

Figs. \ref{fig:plotRAll}(a,b) show the 
zero-temperature, interlayer resistance over a range of Fermi energies
from $\pm 1$ eV around the charge neutrality point in
Fig. \ref{fig:plotRAll}(a) and from $\pm 0.7$ eV in Fig. \ref{fig:plotRAll}(b)
for a range of rotation angles from $0^{\circ}$ to $27.79^{\circ}$.
The lowest curve is the coherent resistance of the ideal AB stack with
$\theta = 0$\degree.
This resistance is the fundamental limiting `quantum resistance' inversely 
proportional to the number of
transverse modes available to carry the current at a given energy.
This quantity has recently been calculated for other materials
to determine the fundamental lower limit on the contact 
resistance.\cite{Rodwell_Lundstro_APL13}
The magnitude of the coherent interlayer resistance
increases several orders of magnitude as the layers become misaligned.
The legends in Figs. \ref{fig:plotRAll}(a,b) are 
ordered according to the size of the corresponding
commensurate unit cell so that, among the rotated interfaces, 
$\theta = 21.8^\circ$ gives the smallest
unit cell and $\theta = 2.87^\circ$ gives the largest unit cell.
For angles $> 7.34^{\circ}$,
the magnitude of the resistance increases with 
the size of the unit cell.
This is the same trend found for the coherent, interlayer resistance
of bilayer graphene
discussed in Ref. [\onlinecite{Avouris_twisted_PRL12}].
The resistances for angles $\leq 7.34^{\circ}$,
fall off rapidly as the energy moves away from the charge neutrality point,
so that at larger energies, this trend fails for the
smaller rotation angles.

All of the angles shown {\em except} $20.31^{\circ}$
fall along the line
of minimum unit-cell size shown in Fig. 2 of Ref. 
[\onlinecite{Shallcross_band_turbo_graphene_PRB10}].
A one to two degree change in the rotation angle can change the
commensurate unit cell size by over 3 orders of magnitude.
Thus, it is interesting to consider two very close angles with 
a large difference in cell size.
Two such angles 
are $21.78^{\circ}$ in Fig. \ref{fig:plotRAll}(a)
and $20.31^{\circ}$ in Fig. \ref{fig:plotRAll}(b).
These two angles differ by $1.47^{\circ}$, yet the
$21.78^{\circ}$ rotation gives the smallest unit cell with a lattice
constant of $6.51$ \AA, and the $20.31^{\circ}$ rotation gives the second largest
unit cell with a lattice constant of $36.23$ \AA.
Near the charge neutrality point, the resistance of the
$20.31^{\circ}$ structure is the highest of all of the structures.
This follows the trend of increasing resistance with unit cell size.
At energies $\pm 0.2$ eV away from the charge neutrality point, the
resistance rapidly falls 4 to 5 orders of magnitude
and approaches the resistance of the $21.78^{\circ}$
structure.
This can be understood by considering the extended zone
scheme of the reciprocal lattice.
At this energy, the Fermi surfaces around the K points 
that coincide in the extended zone scheme
of the $21.78^{\circ}$
structure just begin to touch.
\begin{figure}
\includegraphics[width=3.5in]{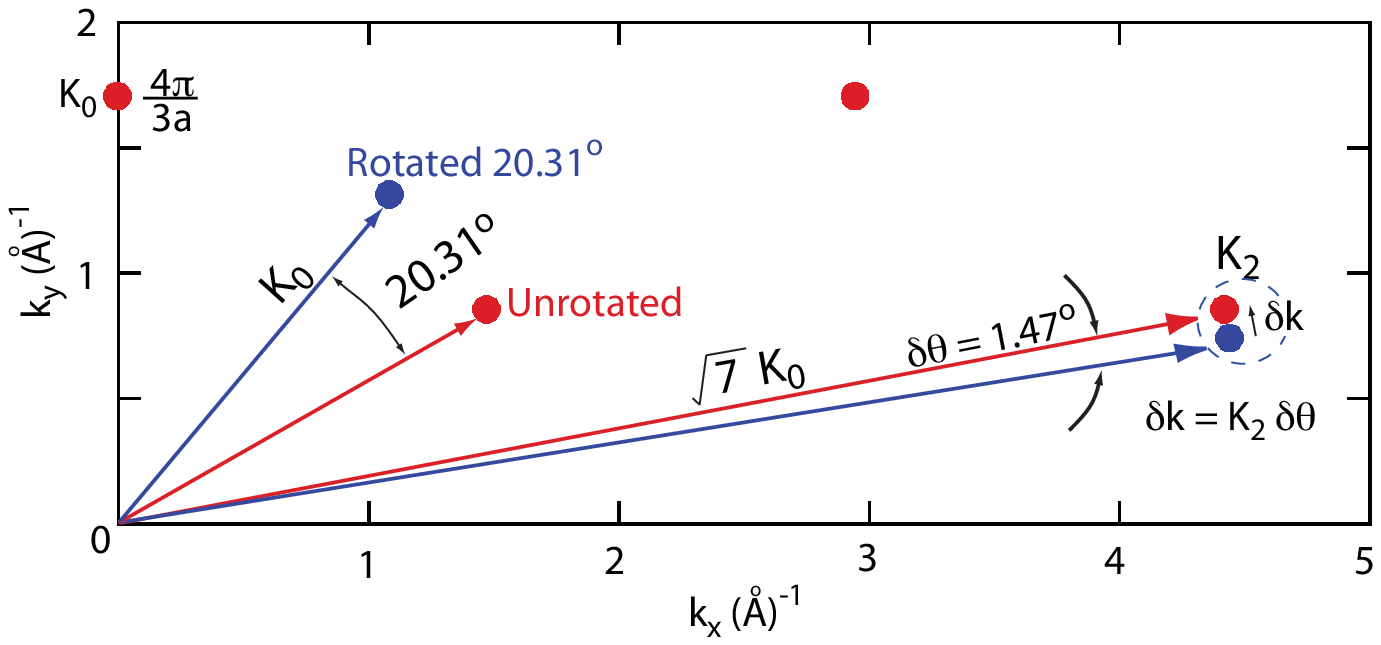}
\caption{
(Color Online) Upper right quadrant of the extended Brillouin zone. 
The unrotated $K$-points are red, and the rotated $K$-points are blue.
The rotation angle is $20.31^{\circ}$.
In the second Brillouin zone, at a distance $\sqrt{7} K_0$ from
$\Gamma$, the $K$-points of the unrotated
and rotated lattices are misaligned by $1.47^{\circ}$.
At low energies, transmission takes place around a $K$ point
with magnitude $\sqrt{217} K_0$. 
At 0.2 eV, transmission begins around the $K$-point at $\sqrt{7} K_0$.
\label{fig:Ext_BZ}
}
\end{figure}

We refer to the magnitude of the K points of the $21.78^{\circ}$ 
structure in the extended zone scheme as $K_2$.
They are illustrated in Fig. \ref{fig:Ext_BZ}.
They lie in the second Brillouin zones and their
magnitude is $K_2 = \sqrt{7} K_0$ where $K_0$ is the magnitude of the
$K$ point in the first Brillouin zone, $K_0 = \frac{4\pi}{3a}$,
and $a = 2.46$ \AA. 
At a rotation angle of $20.31^{\circ}$, these points are
misaligned by $\delta \theta = 1.47^{\circ}$, and their centers
are misaligned in $k$-space by
$\delta k = \sqrt{7} K_0 \delta \theta = 0.116$ \AA$^{-1}$.
The radius $(k_r)$ of the Fermi circle in the $k_x - k_y$ plane
of ideal AB graphite
at $E = 0.2$ eV is $k_r = 0.058 {\rm \AA}^{-1} = \delta k / 2$.
Thus, at $E = 0.2$ eV in the $20.31^{\circ}$ structure, 
the Fermi surfaces begin to touch around the $K_2$ points. 
In the extended zone scheme, at low energies, 
in the $20.31^{\circ}$ structure, 
conduction takes place at $K$ points with a magnitude
of $\sqrt{217} K_0$. 
At $E = 0.2$ eV, conduction begins at
the $K_2$ points with a magnitude of $\sqrt{7} K_0$.
Since the matrix element coupling the states 
decays exponentially with the magnitude of $k$, there is a sudden
decrease in resistance when a channel opens 
at a much smaller $k$-point in the extended 
zone.\cite{Bistritzer_transport_twisted_PRB10} 

As the magnitude of the energy increases, new channels open
around $K$-points in the extended zone.  
When these $K$ points are closer to $\Gamma$, the resistance
suddenly drops and approaches a new value dominated by the
transmission through the smaller $K$ points.
For example, at $\sim 0.8$ eV, the $\theta = 17.89^{\circ}$
structure begins to conduct around the $K_2$ points 
and its resistance falls several orders of magnitude
to that of the $\theta = 21.78^{\circ}$ structure.
When the resistance falls to the same order of magnitude
as that of the ideal AB structure,
it indicates that the transmission is taking place around
the $K$-points of the first Brillouin zone.
At higher energies, this is the dominant transport channel
for all of the low-angle structures as can be seen in
Fig. \ref{fig:plotRAll}(b).

When a new channel opens up in the extended zone scheme,
it will generally appear in the reduced zone of the commensurate
primitive cell as a sudden movement of the transmission in $k$-space.
This results from the fact that new $K$ points in the extended zone do not, 
in general, map onto the $K$ points of the reduced Brillouin zone.
For the $20.31^{\circ}$ structure, the transmission in the reduced
Brillouin zone shifts from the $K$ point to the $M$ point at $E = 0.2$ eV.

The coherent interlayer resistances at $T=300$ K 
are shown in Figs. \ref{fig:plotRAll}(c,d).
They are obtained by convolving the transmission with the room temperature
thermal broadening function in Eq. (\ref{eq:R}) which removes the 
sharpest features from Fig. \ref{fig:plotRAll}(a,b). 
At room temperature, the interlayer resistances for
angles $\leq 7.34^{\circ}$ exponentially
decay with energy torwards the ideal unrotated value.

\begin{figure}
\includegraphics[width=3.5in]{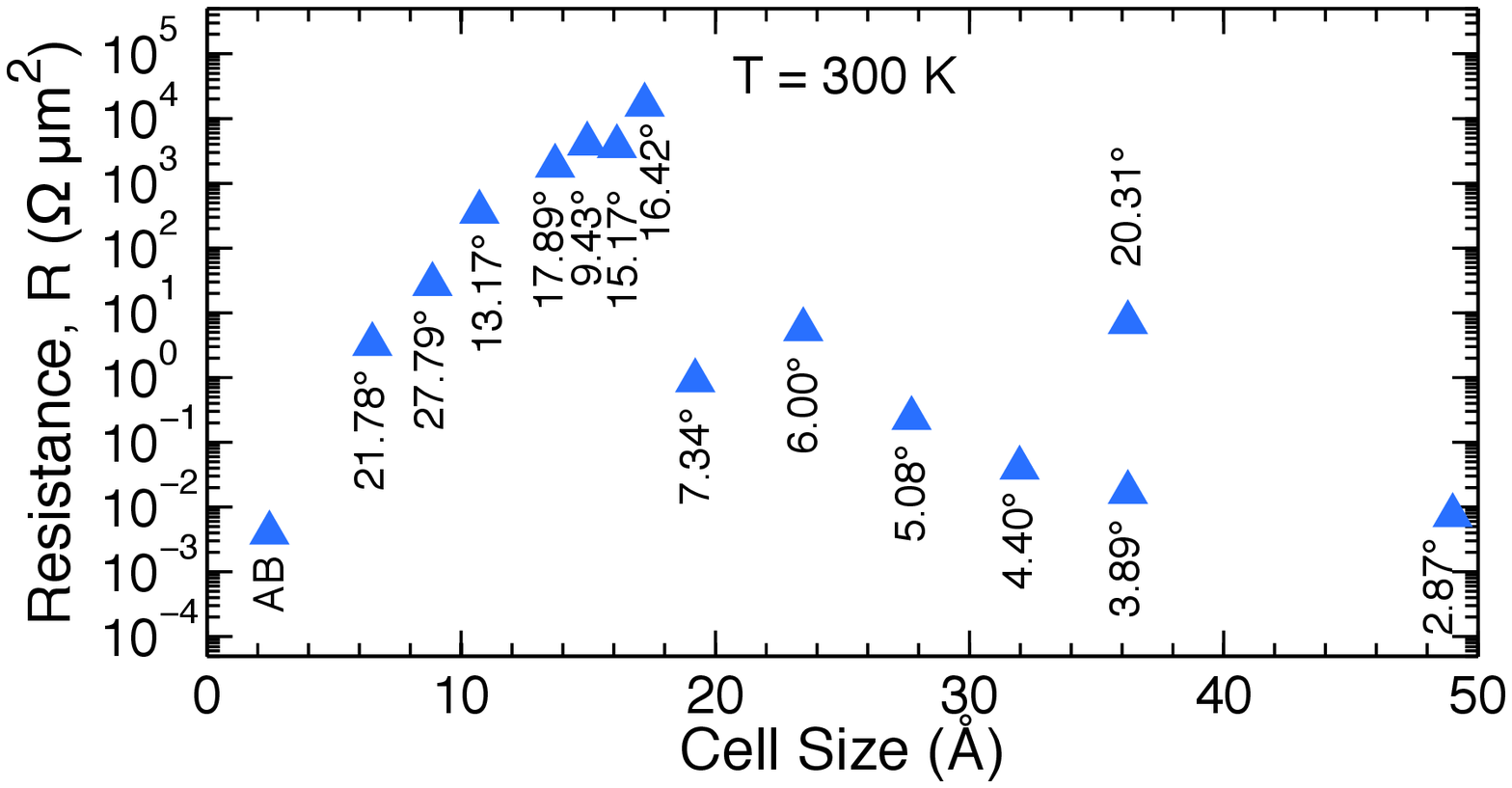}
\caption{
(Color Online) Room temperature coherent ($R_c$) 
resistance as as a function of the commensurate unit cell size 
at $E_F = 0.26$ eV. The corresponding angles are shown in the figure.
\label{fig:plotRtotRoomTempL}
}
\end{figure}

The room temperature resistance values for all structures 
are plotted in Fig. \ref{fig:plotRtotRoomTempL}
at a Fermi energy of 0.26 eV as considered in Refs. 
[\onlinecite{Bistritzer_transport_twisted_PRB10,Avouris_twisted_PRL12}]
for rotated bilayer graphene.
The values are also listed in Table \ref{tab:R}.
The trend of exponentially increasing resistance with unit-cell size is 
clear for rotation angles $\geq 9.34^{\circ}$. 
An abrupt, three order of magnitude discontinuity in the trend
occurs between $9.43^{\circ}$ and $7.34^{\circ}$ for the low rotation angles.
In these structures, transmission is taking place around
the $K$-points of the first Brillouin zone.

There is also the one outlying point from the $20.31^{\circ}$
structure.
While its unit-cell size is huge, and it is not a small angle,
its resistance is far off of the initial trend.
It has the same resistance
as the smallest $21.78^{\circ}$ structure.
That is because, at an energy of 0.26 eV, its
transmission is taking place around the same $K$ points
in the extended zone as that of the $21.78^{\circ}$ structure.

The vast number of huge commensurate primitive cells
a small rotation angle away from much smaller
primitive cells, as shown in Fig. 2 of Ref. 
[\onlinecite{Shallcross_band_turbo_graphene_PRB10}], 
will not follow the exponential trend of resistance versus 
cell size at any energy a few hundred meV away from the charge neutrality point.
Instead, their finite Fermi surfaces will overlap at some much reduced
$K$ point in the extended zone, and those $K$-points, corresponding to a much
smaller cell size, will control the conductance.
In conclusion, the 
quantum-resistance of ideal AB graphene
is on the order of $10^{-3} - 10^{-2}$ $\Omega \mu{\rm m}^2$.
For small misorientation angles, 
the coherent interlayer resistance
exponentially decreases towards the ideal, unrotated 
AB value at higher energies.
For intermediate angles of minimal cell sizes,
the coherent interlayer resistance 
exponentially increases with cell size.
For intermediate angles with very large cell sizes,
the resistance will correspond to a much smaller
cell size of a nearby angle for any finite Fermi energy
of a few hundred meV.
\\
\\
\noindent
{\em Acknowledgement}: This work was supported in part by FAME, 
one of six centers of STARnet, a Semiconductor Research Corporation 
program sponsored by MARCO and DARPA.

\begin{table}
\centering
\begin{tabular}{cccc}
\hline
Rotation Angle $\theta$ & Lattice Constant & No of Atoms & $R_c$ \\
\hline\hline
AB      & 2.46  &  2    & $3.89\times10^{-3}$   \\
\hline
21.78   & 6.51  &  28   & 3.28 \\
\hline
27.79   & 8.87  &  52   & 27.7 \\
\hline
13.17   & 10.72 &  76   &  363 \\
\hline
17.89   & 13.69 &  124  & 1860 \\
\hline
9.43    & 14.96 &  148  & 4080 \\
\hline
15.17   & 16.13 &  172  & 3720 \\
\hline
16.42   & 17.22 &  196  & 16400 \\
\hline
7.34    & 19.21 &  244  & 0.90 \\
\hline
6.00    & 23.46 &  364  & 5.55 \\
\hline
5.08    & 27.71 &  508  & 0.24 \\
\hline
4.40    & 31.97 &  676  & $4.03\times10^{-2}$ \\
\hline
3.89    & 36.23 &  868  & $1.69\times10^{-2}$ \\
\hline
20.31   & 36.23 &  868  & 7.11 \\
\hline
2.87    & 49.01 &  1588 & $7.45\times10^{-3}$ \\
\hline

\end{tabular}
\caption{\label{tab:R}  Coherent resistance $R_c$, 
as a function of rotation angle and primitive-cell lattice constant. 
Resistance units are 
$(\Omega \mu{\rm m}^2)$. 
Angles are in degrees and the lattice constants are in {\AA}.
$T=300$ K and $E_F = 0.26$ eV.
The angles are ordered
according to the supercell size from smallest to largest.
}
\label{tab:Rs}
\end{table}
%


%
%
%

%
%
%
%

%

%
%

%
%

\end{document}